\documentclass[aip, jcp, amsmath, amssymb, reprint]{revtex4-1}

\usepackage{graphicx}
\usepackage{dcolumn}
\usepackage{bm}
\usepackage{color}

\begin{document}


\title{Lattice model of linear telechelic polymer melts. I. Inclusion of chain semiflexibility in the lattice cluster theory}

\author{Wen-Sheng Xu}
\email{wsxu@uchicago.edu}
\affiliation{James Franck Institute, The University of Chicago, Chicago, Illinois 60637, USA}

\author{Karl F. Freed}
\email{freed@uchicago.edu}
\affiliation{James Franck Institute, The University of Chicago, Chicago, Illinois 60637, USA}
\affiliation{Department of Chemistry, The University of Chicago, Chicago, Illinois 60637, USA}

\date{\today}

\begin{abstract}
The lattice cluster theory (LCT) for the thermodynamics of polymer systems has recently been reformulated to treat strongly interacting self-assembling polymers composed of fully flexible linear telechelic chains [J. Dudowicz and K. F. Freed, J. Chem. Phys. \textbf{136}, 064902 (2012)]. Here, we further extend the LCT for linear telechelic polymer melts to include a description of chain semiflexibility, which is treated by introducing a bending energy penalty whenever a pair of consecutive bonds from a single chain lies along orthogonal directions. An analytical expression for the Helmholtz free energy is derived for the model of semiflexible linear telechelic polymer melts. The extension provides a theoretical tool for investigating the influence of chain stiffness on the thermodynamics of self-assembling telechelic polymers, and for further exploring the influence of self-assembly on glass formation in such systems. 
\end{abstract}



\maketitle

\section{Introduction}

Telechelic polymers, containing one bifunctional associative group at each of the chain ends (called ``stickers''), provide a striking example of macromolecules that are capable of supramolecular self-assembly.~\cite{Polymer_49_1425} The self-assembly in telechelics has attracted considerable attention for both technological and fundamental reasons. On one hand, the distinctive characteristics of telechelics, arising from the reversible formation and breakage bonds during the dynamical self-assembly, open the prospect of many new applications,~\cite{POC_7_289, Book_Goodman, Polymer_45_3527, Nature_453_171} which are generally not accessible by conventional polymerization. On the other hand, the theoretical description of the self-assembly process in telechelics represents a large challenge because of the interplay of the strong interactions between the stickers and the van der Waals interactions between the polymer segments. Additional complexity arises from the internal chemical structure of individual telechelic molecules as well as from the additional reversible chain connectivity introduced by the associative clusters.

Theories~\cite{Mac_28_1066, Mac_28_7879, JCP_110_1781, Mac_33_1425, Mac_33_1443, JCP_119_6916, Lan_20_7860, JPSB_45_3285, JCP_131_144906, JPCB_114_12298} and simulations~\cite{Mac_20_1999, JCP_110_6039, EPL_59_384, Polymer_45_3961, JPSB_43_796, PRL_96_187802, PRL_109_238301, JCP_126_044907, JPCM_20_335103, Mac_47_4118, Mac_47_6946} of self-assembly in telechelic polymers traditionally employ highly coarse grained models that represent the assembling molecular species as a structureless entity and hence, neglect the influence of local molecular structure on the self-assembly of telechelic polymers. However, a deep understanding of the relation between molecular structure and physical properties is important in guiding the rational design of telechelic polymer materials. To address this need, Dudowicz and Freed~\cite{JCP_136_064902} have recently developed an intermediate level of coarse grained models that retain minimal aspects of molecular structure and interactions in telechelic polymers. The initial implementation of the model has proceeded by extending the lattice cluster theory (LCT) for the thermodynamics of polymer systems~\cite{JCP_87_7272, Mac_24_5076, ACP_103_335, APS_183_63} to strongly interacting, self-assembling polymers composed of fully flexible linear telechelic chains. The description provided by this extended LCT enables establishing the relation between the molecular structure dependent interaction parameters of the model and the thermodynamic properties of these complex fluids.~\cite{JCP_136_064903, JCP_136_194902} Therefore, the extended LCT for telechelic polymers fills a gap between the limited predictive abilities of the previous theories with structureless monomers and the thermodynamic complexity of the molecular details of the self-assembly processes.

Several improvements are desirable within the LCT for telechelic polymers since the initial study by Dudowicz and Freed~\cite{JCP_136_064902} only considers the simplest case, i.e., fully flexible linear telechelic polymer chains. For example, the applicability of the theory will be enhanced by including the description of monomer structures, a description that has been demonstrated to greatly affect the thermodynamics of polymer systems.~\cite{APS_183_63} Another improvement represents the focus of the present paper, namely, allowing the polymer chains to be semiflexible. Our motivation, in part, comes from the fact that chain semiflexibility plays an important role in determining the self-assembly process and structural properties of telechelic polymers; e.g., computer simulations~\cite{JCP_110_6039, Mac_47_4118, Mac_47_6946} indicate that the structure of the aggregates in telechelic polymers changes significantly with the degree of chain stiffness. More importantly, chain semiflexibility is crucial for exploring glass formation in systems of self-assembling telechelic polymers within the generalized entropy theory (GET),~\cite{ACP_137_125} which is a merger of the LCT for the thermodynamics of semiflexible polymers~\cite{ACP_103_335} and the Adam-Gibbs (AG) relation between the structural relaxation time and the configurational entropy,~\cite{JCP_43_139, JCP_141_141102} because the characteristic glassy behavior does not appear in the GET when the polymer chains are modeled as being fully flexible, as noted in Ref.~\citenum{JCP_140_244905}. Therefore, the present paper further extends the LCT to describe the thermodynamics of a melt of semiflexible linear telechelic chains. The extension in the present paper, thus, also enables investigating the influence of self-assembly on glass formation in telechelic polymer melts in the GET. 

\section{Lattice model of semiflexible linear telechelic polymer melts}

\begin{figure}[tb]
 \centering
 \includegraphics[angle=0,width=0.25\textwidth]{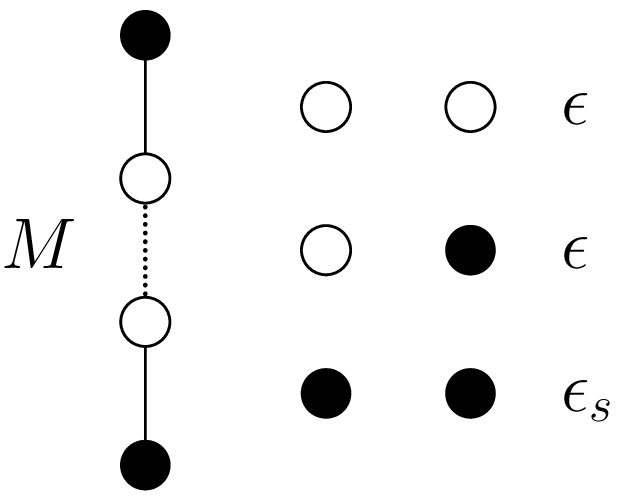}
 \caption{Schematic illustration of the lattice model of a single linear telechelic polymer chain with $M$ united atom groups. Solid circles designate the chain's ends that can participate in sticky interactions. Solid lines represent the bonds between united atom groups, while dotted line indicates the presence of intervening united atom groups and bonds between the two separated portions on the chain. The model prescribes two different nearest neighbor interaction energies $\epsilon$ and $\epsilon_s$ for the weakly interacting interior united atom groups and the stickers, respectively.}
\end{figure}

As in previous work,~\cite{ACP_103_335, JCP_136_064902, JCP_141_044909} the polymer chains are placed on a $d$-dimensional hypercubic lattice with $N_l$ lattice sites, each with $z=2d$ nearest neighbors. The present work considers a compressible (i.e., the lattice contains empty sites) telechelic polymer melt, where the system consists of $m$ semiflexible linear chains with $M$ united atom groups per chain, producing the volume fraction of the polymer chains as $\phi=mM/N_l$. Semiflexibility is treated by introducing a bending energy penalty $E_b$ whenever a pair of consecutive bonds from a single chain lies along orthogonal directions.~\cite{ACP_103_335} Telechelic chains are modeled by distinguishing each chain's end segments (called ``stickers'' and represented as solid circles in Fig. 1) from those lying in the chain interior (depicted by open circles in Fig. 1), as introduced in Ref.~\citenum{JCP_136_064902}. More specifically, two stickers on nearest neighbor lattice sites can form a sticky ``bond''  with an attractive sticky interaction energy $\epsilon_s$ that may greatly exceed the microscopic van der Waals interaction energy $\epsilon$, which describes the interaction strength between two non-stickers as well as between a sticker and a non-sticker (Fig. 1). The stickers are assumed to be mono-functional and hence, each sticker can only participate in one sticky interaction. In accord with the previous work~\cite{ JCP_136_064902, JCP_136_244904} and the analysis of Jacobson and Stockmayer,~\cite{ JCP_18_1600} both cyclic and linear associative clusters may form. Because the lattice model is obviously coarse grained, the sticky energy is generally a temperature dependent parameter, but taken as a constant for simplicity. An attractive microscopic sticky interaction energy $\epsilon_s$ is defined here as negative,\cite{ JCP_136_064902} while the attractive microscopic van der Waals interaction energy $\epsilon$ is treated as positive, in line with the original LCT.~\cite{Mac_24_5076, JCP_87_5534} Since the sticky bonds (which are essentially non-covalent) formed by stickers differ from the covalent bonds within a single chain in real telechelic polymers, semiflexibility constraints are not applied for a pair of sequential bonds containing one sticky bond in the present work, in line with computer simulations of semiflexible telechelic polymers.~\cite{JCP_110_6039, Mac_47_4118, Mac_47_6946} When the sticky interactions are strong, the clusters formed by rigid rod telechelic chains become freely hinged chains that are connected by flexible joints. Therefore, crystallization may be avoided in the models of rigid linear telechelic polymers, a feature that is potentially important for exploring glass formation in such systems.

While the previous study of fully flexible linear telechelic chains with the LCT considers the system as incompressible solutions,~\cite{ JCP_136_064902}
the mathematical equivalence between a compressible melt and an incompressible solution allows drawing conclusions for both types of systems. The model for an incompressible solution replaces the empty sites by one-bead solvent molecules instead. The excess free energy of a compressible polymer melt is isomorphic to that of an incompressible polymer solution, with the $\epsilon$ parameter being replaced by the exchange energy $\epsilon_{\text{ex}}=\epsilon_{pp}+\epsilon_{ss}-2\epsilon_{ps}$, where $\epsilon_{pp}$, $\epsilon_{ss}$ and $\epsilon_{ps}$ represent the strengths of the nearest neighbor interaction between two polymer segments, two solvent molecules and a polymer segment and a solvent molecule, respectively. Because the theory in the present paper will be used in a subsequent paper to explore glass formation in linear telechelic polymer melts, the discussion of the model and results are expressed in the present work with reference to compressible melts.

\section{Lattice cluster theory for semiflexible linear telechelic polymer melts}

As demonstrated in a previous paper,~\cite{ JCP_136_064902} the development of the LCT for associating polymers begins by noting that the specific Helmholtz free energy $f=F/N_l$ (with $F$ the total Helmholtz free energy ) for a self-assembling system is the sum of the specific free energy $f_o$ in the absence of strong sticky interactions and a contribution $f_s$ arising from the sticky interactions,
\begin{equation}
f=f_o+f_s.
\end{equation}
Specializing to the model in the present work, $f_o$ is the specific free energy for a polymer melt of semiflexible linear chains, where each chain's ends are indistinguishable from the other united atom groups. This is our reference system. The free energy has been derived for the model of multicomponent systems with structured monomer chains in the previous paper~\cite{ACP_103_335} with some corrections given in Ref.~\citenum{JCP_141_044909}. Therefore, this section first summarizes the result for the free energy $f_o$ specialized to the model of polymer melts composed of semiflexible linear chains, and then describes the evaluation of $f_s$ with the emphasis on how to include chain semiflexibility in the presence of sticky interactions.

\subsection{Free energy for the reference system}

The LCT yields the specific Helmholtz free energy $f_o$ of a semiflexible polymer melt in the general form,~\cite{ACP_103_335, JCP_141_044909}
\begin{equation}
\beta f_o=\beta f_o^{mf}-\sum_{i=1}^6C_i\phi^i,
\end{equation}
where $\beta=1/(k_BT)$ with $k_B$ being Boltzmann's constant and $T$ designating the absolute temperature. The term $\beta f_o^{mf}$ represents the zeroth-order mean-field contribution, and has the following form for a melt of semiflexible linear chains,
\begin{eqnarray}
\beta f_o^{mf}=&&\frac{\phi}{M}\ln\left(\frac{2\phi}{zM}\right)+\phi\left(1-\frac{1}{M}\right)+ (1-\phi)\ln(1-\phi)\nonumber\\
&&
-\phi \frac{N_{2}}{M}\ln(z_b),
\end{eqnarray}
where $N_2$ is the number of runs of two consecutive bonds in a single chain, and $z_b=(z_p-1)\exp(-\beta E_b)+1$ with $z_p=z/2$ and $E_b$ being the bending energy. The coefficients $C_i$ $(i=1, ..., 6)$ are obtained by collecting terms corresponding to a given power of $\phi$, and these coefficients are generally a function of $z$, $T$, $\epsilon$, $E_b$, and a set of counting indices $u_i=N_i/M$ $(i=1, ..., 4)$, where the counting factor $N_i$ denotes the number of runs of $i$ consecutive bonds in a single chain and simplifies to $N_i=M-i$ for linear chains. The explicit expressions for $C_i$ $(i=1, ..., 6)$ for a melt of semiflexible linear chains are provided in Appendix A.

\subsection{LCT partition function for semiflexible linear telechelic polymer melts and its diagrammatic representation}

As explained in Refs.~\citenum{JCP_136_064902, JCP_130_061103}, the main idea behind the extension of the LCT to strongly interacting polymer systems lies in the extraction of terms associated with the strong interactions from the cluster expansion. Consequently, the partition function $W$ for the self-assembling systems is first defined for systems with a constant number $H$ of sticky bonds and then is summed over all possible $H$ to yield $W$.

Imposing the constraints associated with the presence of excluded volume interactions, chain connectivity, semiflexibility, nearest neighbor van der Waals interactions, and sticky interactions, the LCT partition function $W(H,N_s,m,M)$ for a melt of semiflexible linear telechelic chains is derived as
\begin{eqnarray}
W=&&\frac{1}{2^mm!}\sum_{\{\mathbf{r_i}\}}{\textstyle{'}}
\left\{ \prod_{u=1}^{m}  \sum_{\mu=1}^{z} \left[ \prod_{i=1}^{M-1}\delta(\mathbf{r}_{u,i}, \mathbf{r}_{u,i+1}+\mathbf{a}_{\mu}) \right. \right. \nonumber\\
&&
\times \prod_{j=1}^{M-2}[E+(1-E)\delta(\mu_{u,j}, \mu_{u,j+1})] \nonumber\\
&&
\times \left. \prod_{v=1}^{m} \prod_{k=1}^{M-1}\left(1+f_{\text{pair}}\sum_{\nu=1}^{z}\delta(\mathbf{r}_{u,i}, \mathbf{r}_{v,k}+\mathbf{a}_{\nu})\right) \right] \nonumber\\
&&
\times \frac{N_s!}{(N_s-2H)!H!2^H} \exp(-\beta H \epsilon_s) \nonumber\\
&&
\times \left. \prod_{t=1}^{H}\left[\sum_{\gamma=1}^{z}\delta(\mathbf{r}_{i_p}, \mathbf{r}_{i_q}+\mathbf{a}_{\gamma})\right] \right\},
\end{eqnarray}
where the symbol $\sum'_{\{\mathbf{r_i}\}}$ represents a restricted sum over the positions of all united atom groups subject to the constraint prohibiting multiple occupancy of any lattice site, $\delta$ is the Kronecker delta function (the physical meaning of each specific $\delta$ is explained in detail in Refs.~\citenum{ACP_103_335, JCP_136_064902}), $\mathbf{a}$ is the lattice vector, the factor $E$ is defined by $E=\exp(-\beta E_b)$, $f_{\text{pair}}=\exp(\beta\epsilon)-1$ is the Mayer $f$-function, $N_s$ is the total number of stickers in the system, $\epsilon_s$ is the sticker-sticker interaction energy, the index $t$ labels the sticky bonds, and $\mathbf{r}_{i_p}$ and $\mathbf{r}_{i_q}$ are the positions of the nearest neighbor stickers $p$ and $q$, which belong to the set $\{\mathbf{r}_{i}\}$ of lattice sites that are occupied by the polymer chains. The factors $2^{-m}$ and $1/m!$ in Eq. (4) account for the indistinguishability of the two chain ends and of the chains themselves, while the combinatorial factor $N_s!/[(N_s-2H)!H!2^H]$ in Eq. (4) accounts for the number of ways of forming a total of $H$ sticky bonds from the total number $N_s$ of stickers in the system.

In analogy to the treatment of the partition function for the reference system,~\cite{ACP_103_335} the partition function $W(H,N_s,m,M)$ for the telechelic polymers can schematically be rewritten in the convenient form of multiple cluster expansions,
\begin{eqnarray}
W=&&\sum_{\{\mathbf{r}\}}{\textstyle{'}}\frac{W_{o}^{mf}W_{s}^{mf}(N_l-mM)!}{N_{l}!}
\left[ \prod_{\text{bond}}\left( 1+X_{\text{bond}} \right) \right.\nonumber\\
&&
\times \prod_{\text{pair}} \left( 1+GX_{\text{pair}} \right) \times \prod_{\text{bend}}\left( 1+KY_{\text{bend}} \right) \nonumber\\
&&
\times \left. \prod_{\text{sticky}}\left( 1+X_{\text{sticky}} \right) \right],
\end{eqnarray}
where the zeroth-order mean-field terms $W_{o}^{mf}$ and $W_{s}^{mf}$ are provided by Eq. (19) in Ref.~\citenum{ACP_103_335} and Eq. (11) in Ref.~\citenum{JCP_136_064902}, respectively. The terms $X_{\text{bond}}$, $G$, $X_{\text{pair}}$, $K$, $Y_{\text{bend}}$, and $X_{\text{sticky}}$ in Eq. (11) are given by
\begin{equation}
X_{\text{bond}}=\frac{N_l}{z}\left[ \sum_{\mu=1}^{z}\delta(\mathbf{r}_{u,i},\mathbf{r}_{u,i+1}+\mathbf{a}_{\mu})-\frac{z}{N_l} \right],
\end{equation}
\begin{equation}
G=\frac{zf_{\text{pair}}/N_l}{1+zf_{\text{pair}}/N_l},
\end{equation}
\begin{equation}
X_{\text{pair}}=\frac{N_l}{z}\left[ \sum_{\nu=1}^{z}\delta(\mathbf{r}_{u,i},\mathbf{r}_{v,k}+\mathbf{a}_{\nu})-\frac{z}{N_l} \right],
\end{equation}
\begin{equation}
K=\frac{1-E}{(z_p-1)E+1}, 
\end{equation}
\begin{equation}
Y_{\text{bend}}=z\delta(\mu_{u,j},\mu_{u,j+1})-1,
\end{equation}
and
\begin{equation}
X_{\text{sticky}}=\frac{N_l}{z}\left[ \sum_{\gamma=1}^{z}\delta(\mathbf{r}_{i_p},\mathbf{r}_{i_q}+\mathbf{a}_{\gamma})-\frac{z}{N_l} \right].
\end{equation}
Equations (11-17) are achieved by taking advantage of the fact that each Kronecker delta function in Eq. (4) can be expressed in the convenient form,
\begin{equation}
\delta=A+(\delta-A)=A\left( 1+\frac{\delta-A}{A} \right),
\end{equation}
where the term $A$ represents an average of the contribution from $\delta$ and thus defines a factor appearing in the zeroth-order mean-field term $W_o^{mf}$ or $W_s^{mf}$.

\begin{figure}[tb]
 \centering
 \includegraphics[angle=0,width=0.48\textwidth]{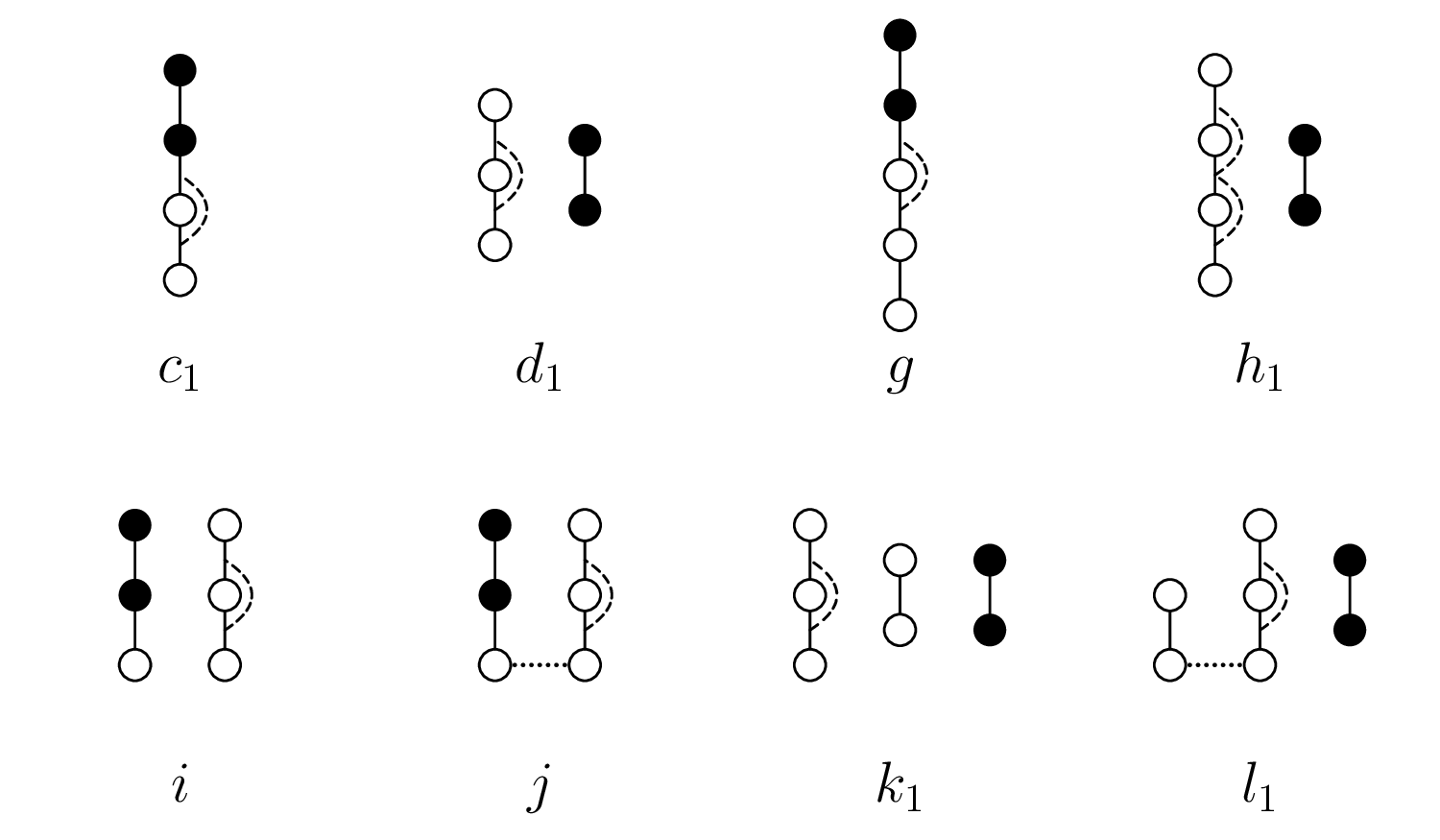}
 \caption{Contributing cumulant diagrams with a single sticky bond for the model of semiflexible linear telechelic polymer melts. Each diagram contains at least one bending constraint line but no interaction lines. Open and solid circles correspond to non-sticky and sticky united atom groups, respectively. Solid straight lines denote non-sticky or sticky bonds, while dotted straight lines in $j$ and $l_1$ indicate the presence of one or more intervening bonds within a chain between the depicted bonds. The diagrams are derived from Fig. 5 in Ref.~\citenum{JCP_136_064902}.}
\end{figure}

\begin{figure}[tb]
 \centering
 \includegraphics[angle=0,width=0.10\textwidth]{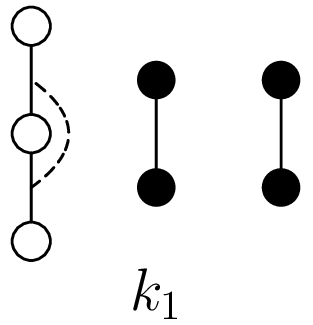}
 \caption{Contributing cumulant diagram with two sticky bonds, one bending constraint line, but no interaction lines for the model of semiflexible linear telechelic polymer melts. The diagram is derived from Fig. 6 in Ref.~\citenum{JCP_136_064902}.}
\end{figure}

\begin{figure}[tb]
 \centering
 \includegraphics[angle=0,width=0.48\textwidth]{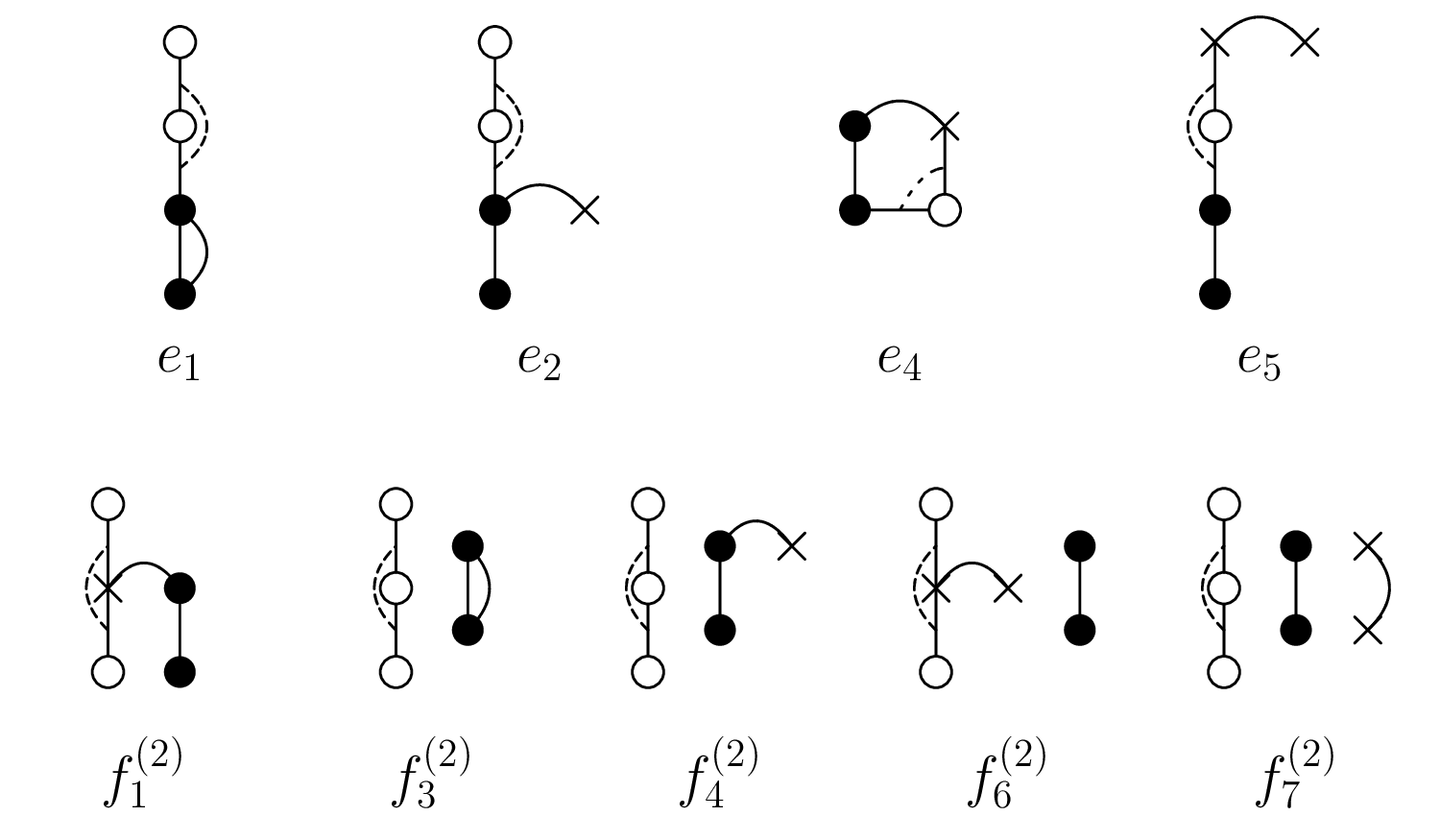}
 \caption{Contributing cumulant diagrams with a single sticky bond for the model of semiflexible linear telechelic polymer melts. Each diagram contains one bending constraint line and one interaction line. The diagrams are derived from Fig. 8 in Ref.~\citenum{JCP_136_064902}.}
\end{figure}

Expanding the product in Eq. (5) leads naturally to the cluster expansion for the partition function $W(H,N_s,m,M)$,
\begin{eqnarray}
W=&&W_o^{mf}W_s^{mf}\left[ 1+\frac{(N_l-mM)!}{N_{l}!}\sum_{\{\mathbf{r}\}}{\textstyle{'}} \left(\sum_{i}t_{i}
\right. \right. \nonumber\\
&&
\left. \left. +\sum_{i,j}t_{i}t_{j}+\sum_{i,j,k}t_{i}t_{j}t_{k}+\cdots\right) \right],
\end{eqnarray}
where $t_{\alpha}$ $(\alpha=i,j,k,...)$ is either $X_{\text{bond}}$, $GX_{\text{pair}}$, $KY_{\text{bend}}$, or $X_{\text{sticky}}$. The systematic calculation of contributions from the cluster corrections in Eq. (13) can be transformed into explicit expressions for evaluating contributions by representing each $t_{\alpha}$ factor diagrammatically.~\cite{ACP_103_335, JCP_87_7272}  Specifically, the diagrammatic representation depicts a bond (which is either a non-sticky or a sticky bond, corresponding to the factor $X_{\text{bond}}$ or $X_{\text{sticky}}$) by a solid straight line, an interaction between two nearest neighbor united atom groups (i.e., the factor $GX_{\text{pair}}$) by a solid curved line, and a bending constraint (i.e., the factor $KY_{\text{bend}}$) by a dashed curved line. Moreover, non-stickers and stickers are pictured by crosses and solid circles when they participate in weak van der Waals interactions. Equation (13) can then be rendered more compact using this diagrammatic representation,
\begin{eqnarray}
W=&&W_o^{mf}W_s^{mf}\nonumber\\
&&
\times \left[ 1+\sum_{B, l, b, S}\gamma_{D}(B, l, b, S)D(B, l, b, S) \right], 
\end{eqnarray}
where $\gamma_{D}(B, l, b, S)$ and $D(B, l, b, S)$ are the combinatorial and connectivity factors, respectively, for a given diagram with $B$ non-sticky bonds, $l$ interaction lines, $b$ bending constraint lines, and $S$ sticky bonds. The contributions to the partition function from the sticky interactions that are retained originate from the zeroth-order mean-field term $W_s^{mf}$ and those diagrams with one or more sticky bonds that are consistent with the current LCT retention of all diagrams with $B+l+S\leq4$ bonds and interaction lines. Dudowicz and Freed~\cite{JCP_136_064902} derived all relevant diagrams with sticky bonds for fully flexible linear telechelic polymers. The diagrams including the bending constraint lines can then be constructed by adding dashed curved lines, connecting pairs of consecutive non-sticky bond correlation lines in the relevant diagrams. Figures 2-4 provide all diagrams with $S>0$ necessary for including the description of chain semiflexibility within the LCT for linear telechelic polymer melts.

\subsection{Evaluation of the diagrams with semiflexible constraints and sticky bonds}

We briefly describe the evaluation for semiflexible, sticky diagrams of $\gamma_{D}(B, l, b, S)$ and $D(B, l, b, S)$, the factors that are necessary to calculate the diagrams. The combinatorial factor $\gamma_{D}(B, l, b)$ has the form,
 \begin{eqnarray}
\gamma_{D}(B, l, b, S)=s_D\gamma(B, l, b, S),
\end{eqnarray}
where $s_D$ is the symmetry number and $\gamma(B, l, b, S)$ is the number of ways of selecting the set of non-sticky bonds, sticky bonds, and uncorrelated interacting united atom groups from all polymer chains in the system. The symmetry number $s_D$ for a diagram with $S$ sticky and $B$ non-sticky bonds is identical to that for the corresponding diagram with $S+B$ non-sticky bonds.~\cite{JCP_136_064902} However, as illustrated in the example below, certain restrictions arise in evaluating $\gamma(B, l, b, S)$ for diagrams with $S>0$ due to the presence of sticky bonds.~\cite{JCP_136_064902}

The connectivity factor $D(B, l, b, S)$ for a diagram with $S>0$ is identical to that for the corresponding diagram where the sticky bonds are replaced by the non-sticky bonds,~\cite{JCP_136_064902} and hence, the previous method can directly be used to compute $D(B, l, b, S)$.~\cite{ACP_103_335, JCP_141_044909} The connectivity factor $D(B, l, b, S)$ can be written as
\begin{eqnarray}
D(B, l, b, S)=\frac{d(B, l, b, S)}{\alpha}G^{l}K^{b},
\end{eqnarray}
where $\alpha=N_l(N_l-1)\cdots(N_l-N_v+1)$ with $N_v$ being the number of united atom groups in the given diagram. The factor $d(B, l, b, S)$ depends on the lattice structure and can be calculated in terms of contracted diagrams,~\cite {JCP_87_7272} a set of new diagrams that are obtained by merging sets of two or more vertices for the given diagram. Specifically, $d(B, l, b, S)$ is evaluated by
\begin{eqnarray}
d(B, l, b, S)=\sum_{c}f_{B,c}R_{B,c},
\end{eqnarray}
where $c$ is a sequential counting index and the coefficient $f_{B,c}$ is the product of a contraction factor $\prod_{\lambda=1}^{N'_v}(-1)^{k_{\lambda}-1}(k_{\lambda}-1)$ and the number of ways of forming a contracted diagram with $N'_v$ vertices by merging the distinguishable vertices $k_1, k_2, ..., k_{N'_v}$ in the original uncontracted diagram. The expressions for the relevant contracted diagrams $R_{B,c}$ have been provided in Fig. 2 of Ref.~\citenum{JCP_141_044909}. 

\begin{figure}[tb]
 \centering
 \includegraphics[angle=0,width=0.45\textwidth]{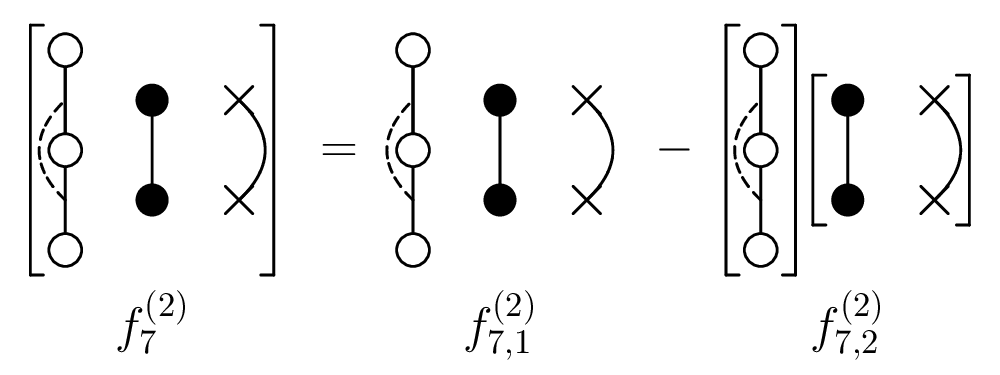}
 \caption{Illustration of the evaluation of the diagram $f_7^{(2)}$ in Fig. 4. The cumulant diagram $f_7^{(2)}$ is composed of two individual diagrams $f_{7,1}^{(2)}$ and $f_{7,2}^{(2)}$, whose combination yields the final result for $f_7^{(2)}$.}
\end{figure}

\begin{table*}[tb]
\caption{Expressions for the terms that are necessary for the evaluation of each individual diagram and hence the cumulant diagram $f_{7}^{(2)}$ in Fig. 4.}
\begin{ruledtabular}
\begin{tabular}{lccc}
\textrm{Diagram}&
\textrm{$\alpha$}&
\textrm{$d$}&
\textrm{$\gamma_D$}\\
\colrule
$f_{7,1}^{(2)}$ & $\prod_{i=0}^{6}(N_l-i)$ & $-2R_{4,2}-8R_{4,5}+16R_{4,10}+32R_{4,11}$ & $H(mN_2-2N_{2e})(mM-5)(mM-6)/2!$\\
$f_{7,2}^{(2)}$ & $\prod_{i=0}^{2}(N_l-i)\prod_{i=0}^{3}(N_l-i)$ & $-2R_{4,2}$ & $H(mN_2)(mM-2)(mM-3)/2!$\\
\end{tabular}
\end{ruledtabular}
\end{table*}

The diagrammatic representation of the Helmholtz free energy $\beta F=-\ln W$ arises from expanding $\ln W$ in a Taylor series and from collecting the resulting contributions into cumulant diagrams.~\cite{JCP_87_7272} Additionally, the Mayer $f$-function is treated using a high temperature expansion.~\cite{ACP_103_335} The diagram $f_{7}^{(2)}$ in Fig. 4 is selected to illustrate the evaluation process for diagrams with sticky bonds, and to further clarify the individual contributions to $\gamma_{D}(B, l, b, S)$ and $D(B, l, b, S)$ and the concept of cumulant diagrams.

The cumulant diagram $f_{7}^{(2)}$ is a combination of two individual diagrams, as depicted in Fig. 5. The topology of each individual diagram can only be constructed in one possible manner by removing the solid curved interaction line and labeling all united atom groups as distinguishable. Therefore, the symmetry number is $s_D=1$ for both individual diagrams. Table I provides a summary for each term of the other factors in $\gamma_{D}(B, l, b, S)$ and $D(B, l, b, S)$. The individual diagram $f_{7,1}^{(2)}$ is used as an example to explain the meaning of each term. The bond topology delineated by the diagram $f_{7,1}^{(2)}$ means that neither of the two consecutive non-sticky bonds can be a bond sequential to the sticky bond. This restriction leads to the factor $H(mN_2-2N_{2e})(mM-5)(mM-6)/2!$, which is the number of ways of selecting the combination of one sticky bond, two consecutive non-sticky bonds, and uncorrelated interacting united atom groups from all polymer chains in the system. The symbol $N_{2e}$ is defined as equal to half of the number of runs of two consecutive bonds in a single chain for which one of the bonds links a sticker with a non-sticker. Hence, $N_{2e}=1$ for linear chains. The diagram $f_{7,1}^k$ contains seven united atom groups, and thus the factor $\alpha$ simply reads $\alpha=\prod_{i=1}^{6}(N_l-i)$. The factor $d=-2R_{4,2}-8R_{4,5}+16R_{4,10}+32R_{4,11}$ with the expressions of $R_{B,c}$ given in Fig. 2 of Ref.~\citenum{JCP_141_044909} is obtained from the contracted diagrams generated when both the interaction line and the sticky bond are replaced by non-sticky bonds. Combining the results from both individual diagrams according to the rules given in Fig. 5 leads to the contribution to the free energy from the cumulant diagram $f_{7}^{(2)}$ as
\begin{eqnarray}
f_7^{(2)}\rightarrow\frac{N_l}{z}Ky\epsilon(4u_2\phi^3-6u_2\phi^2-2N_{2e}\phi^2),
\end{eqnarray}
where the variable $y=H/N_l$ arises from the factor $\gamma_{D}(B, l, b, S)$.

\subsection{Contributions to the free energy arising from the sticky interactions}

Combining all contributions arising from the sticky interactions [i.e., the zeroth-order mean-field term $\beta f_s^{mf}=-1/N_l\ln W_s^{mf}$ and the diagrams with $S>0$ in Eq. (20)] leads to the final expression for $f_s$,
\begin{equation}
\beta f_s=\beta f_s^{mf}-\sum_{i=1}^4Y_iy^i.
\end{equation}
The zeroth-order mean-field term $\beta f_s^{mf}$ reads~\cite{JCP_136_064902}
\begin{eqnarray}
\beta f_s^{mf}=&&-\phi x\ln(\phi x)+(\phi x-2y)\ln(\phi x-2y)\nonumber\\
&&
+y\left[1+\ln\left(\frac{2y}{z}\right)+\beta\epsilon_s\right],
\end{eqnarray}
where $x=2/M$ is the fraction of stickers in a single chain. The coefficients $Y_i$ $(i=1,...,4)$ are obtained by collecting terms corresponding to a given power of $y$ and can be organized in powers of the polymer volume fraction $\phi$,
\begin{equation}
Y_i=\sum_{j=0}^{j_{max}}Y_{i,j}\phi^j,
\end{equation}
where $j_{max}=5$, $4$, $2$, and $0$ for $i=1$, $2$, $3$, and $4$, respectively, and the explicit expressions for $Y_{i,j}$ are provided in Appendix B.

The variable $y$ in Eqs. (19) and (20) is determined by the maximum term method, i.e., by applying the condition,~\cite{JCP_136_064902}
\begin{eqnarray}
\left. \frac{\partial (\beta f_s)}{\partial y}\right|_{N_l, T, \phi}=0,
\end{eqnarray}
which, in turn, takes the form,
\begin{eqnarray}
(\phi x-2y)^2-\left(\frac{2y}{z}\right)\exp\left(\beta\epsilon_s-\sum_{i=1}^4iY_iy^{i-1}\right)=0.
\end{eqnarray}
Equation (23) must be solved numerically, producing the concentration $y^{\ast}$ of the sticky bonds that, when substituted into Eqs. (19) and (20), determines the contributions to the free energy $f_s$ arising from the sticky interactions. Notice that $y^{\ast}$ depends on all molecular and thermodynamic parameters including $T$, $\phi$, $M$, $\epsilon$, $E_b$, and $\epsilon_s$. Because the chain's ends are assumed to be mono-functional in the present model, the volume fraction of the active stickers (i.e., those participating in sticky interactions) is simply $2y^{\ast}$, while the upper limit for $y^{\ast}$ is just $y_{max}^{\ast}=\phi/M$. While the current version of the LCT provides no information concerning the concentration of sticky bonds in the cyclic clusters, cyclic clusters may form, as introduced in Sec. II.

\begin{table}[tb]
	\caption{Summary of the meanings of the parameters that appear in the free energy $f$ for semiflexible linear telechelic melts.}
	\begin{ruledtabular}
		\begin{tabular}{lc}
			\textrm{Symbol}&
			\textrm{Meaning}\\
			\colrule
			$d$ or $z/2$ & spatial dimension \\
			$T$ & absolute temperature \\
			$\phi$ & polymer volume fraction \\
			$M$ & molecular weight \\
			$u_i$ $(i=1, ..., 4)$ & counting indices \\
			$y^{\ast}$ & concentration of the sticky bonds \\
			$\epsilon$ & van der Waals interaction energy \\
			$\epsilon_s$ & sticky interaction energy \\
			$E_b$ & bending energy \\
		\end{tabular}
	\end{ruledtabular}
\end{table}

Finally, the specific free energy $f$ for a melt of semiflexible linear telechelic chains appears as
\begin{eqnarray}
\beta f=&&\beta f_o-\phi x\ln(\phi x)+(\phi x-2y^{\ast})\ln(\phi x-2y^{\ast})\nonumber\\
&&
+y^{\ast}\left[1+\ln\left(\frac{2y^{\ast}}{z}\right)+\beta\epsilon_s\right]-\sum_{i=1}^4Y_i(y^{\ast})^i,
\end{eqnarray}
where the free energy $f_o$ of the reference system is given in Subsection III A. For convenience in understanding and usage of the theory, Table II summarizes the meanings of the parameters that appear in $f$.

\section{Discussion}

The increasing scientific interest in telechelics and their technological importance has motivated a number of theoretical investigations on the self-assembly of telechelic polymers. While theories of self-assembly in telechelic polymers traditionally employ highly coarse grained models that represent the assembling molecular species as a structureless entity,~\cite{Mac_28_1066, Mac_28_7879, JCP_110_1781, Mac_33_1425, Mac_33_1443, JCP_119_6916, Lan_20_7860, JPSB_45_3285, JCP_131_144906, JPCB_114_12298} the LCT of Dudowicz and Freed~\cite{JCP_136_064902} aims at providing a deep understanding of the relation between molecular structure and thermodynamic properties in telechelic polymers. Therefore, the LCT offers the possibility to yield a framework for extracting and organizing information that is useful for designing telechelic polymer materials. 

The initial study by Dudowicz and Freed~\cite{JCP_136_064902} considers for simplicity the fully flexible linear polymers. However, chain stiffness is known to significantly influence the thermodynamics of polymer systems and hence, the role of chain semiflexibility in the self-assembly of telechelic polymers remains to be investigated within the LCT. Moreover, the model of fully flexible linear telechelic polymers is not suitable for exploring glass formation within the GET~\cite{ACP_137_125} because the characteristic glassy behavior appears in the GET only if the polymer chains are modeled as being semiflexible. To address the above needs, we further extend the LCT for the thermodynamics of linear telechelic polymer melts to include a description of chain semiflexibility. Following the previous treatment of chain semiflexibility within the LCT,~\cite{ACP_103_335} chain semiflexibility is treated by introducing a bending energy penalty whenever a pair of consecutive bonds from a single chain lies along orthogonal directions. We provide an analytical expression for the Helmholtz free energy for the model of semiflexible linear telechelic polymer melts. 

The present extension provides a theoretical tool for investigating the influence of chain stiffness on the thermodynamic properties of self-assembling telechelic polymers.  By combining the present extension of the LCT with the AG relation,~\cite{JCP_43_139, JCP_141_141102} we can provide a similar generalization of the GET of polymer glass formation,~\cite{ACP_137_125} thereby permitting exploring the influence of self-assembly on glass formation in telechelic polymers. Because of the great algebraic complexity of the extension, computations for the influence of chain stiffness on the basic thermodynamic properties of self-assembling telechelic polymers will be provided in paper II of this series.~\cite{Paper2}

\begin{acknowledgments}
We thank Jacek Dudowicz for help on the calculations of diagrams with sticky bonds. This work is supported by the National Science Foundation (NSF) Grant No. CHE-1363012.
\end{acknowledgments}

\appendix

\section{Summary of the coefficients that appear in the free energy for the reference system}

The explicit expressions for the coefficients that appear in the free energy $f_o$ for a melt of semiflexible linear chains are provided in the following:

\begin{eqnarray}
	C_1=&&-\frac{u_2g}{z}-\frac{u_2g^2/2-u_4(1-3g+2g^2-g^3)}{z^2}\nonumber\\
	&&
	+\left(u_1+\frac{u_3g^2}{z}\right)(\beta\epsilon),
\end{eqnarray}

\begin{eqnarray}
	C_2=&&\frac{u_1^2}{z}+\frac{u_2^2g^2+2u_1u_3g^2}{z^2}\nonumber\\
	&&
	+\left(z/2-2u_1+\frac{2u_1^2-2u_3g^2-4u_1u_2g}{z}\right)(\beta\epsilon)\nonumber\\
	&&
	+(z/4-u_1+u_2g+u_1^2)(\beta\epsilon)^2,
\end{eqnarray}

\begin{eqnarray}
	C_3=&&\frac{2u_1^3/3-4u_1^2u_2g}{z^2}\nonumber\\
	&&
	+\left(u_1+\frac{u_3g^2-4u_1^2+8u_1u_2g+4u_1^3}{z}\right)(\beta\epsilon)\nonumber\\
	&&
	+(-z/2+7u_1/2-3u_2g-6u_1^2)(\beta\epsilon)^2,
\end{eqnarray}

\begin{eqnarray}
	C_4=&&\frac{2u_1^4}{z^2}+\left(\frac{2u_1^2-8u_1^3-4u_1u_2g}{z}\right)(\beta\epsilon)\nonumber\\
	&&
	+(z/4-4u_1+3u_2g+12u_1^2)(\beta\epsilon)^2,
\end{eqnarray}

\begin{eqnarray}
	C_5=&&\left(\frac{4u_1^3}{z}\right)(\beta\epsilon)+(3u_1/2-u_2g-10u_1^2)(\beta\epsilon)^2,
\end{eqnarray}

\begin{eqnarray}
	C_6=(3u_1^2)(\beta\epsilon)^2,
\end{eqnarray}

where $g=z_p\exp(-\beta E_b)/z_b$ is called the bending energy factor, $\epsilon$ is the nearest neighbor van der Waals interaction energy, and the counting indices $u_i$ $(i=1, ..., 4)$ designate the ratio $N_i/M$ with $N_i$ being the number of runs of $i$ sequential bonds in a single chain. The relation between $N_i$ and $M$ simplifies to $N_i=M-i$ for linear chains. 

\section{Summary of the coefficients that appear in contributions to the free energy arising from the sticky interactions}

The explicit expressions for the coefficients that appear in contributions $f_s$ to the free energy arising from the sticky interactions for a melt of semiflexible linear telechelic chains are provided in the following:

\begin{subequations}
	\begin{eqnarray}
		Y_{1,0}=&&-\frac{2}{z}-\frac{1+2N_{2e}g-N_{3e}(2-4g)}{z^2}\nonumber\\
		&&
		+\left(1+\frac{2N_{2e}g}{z}\right)(\beta\epsilon),
	\end{eqnarray}
	
	\begin{eqnarray}
		Y_{1,1}=&&\frac{2u_1}{z}-\frac{2u_1-4u_1N_{2e}g-4u_2g-2u_3g^2}{z^2}\nonumber\\
		&&
		-\left(2+\frac{4u_1+4u_2g+4N_{2e}g}{z}\right)(\beta\epsilon)\nonumber\\
		&&
		+(1+2u_1)(\beta\epsilon)^2,
	\end{eqnarray}
	
	\begin{eqnarray}
		Y_{1,2}=&&-\frac{6u_1^2+8u_1u_2g}{z}\nonumber\\
		&&
		+\left(1+\frac{8u_1+8u_2g+2N_{2e}g+12u_1^2}{z}\right)(\beta\epsilon)\nonumber\\
		&&
		-(5/2+12u_1)(\beta\epsilon)^2,
	\end{eqnarray}
	
	\begin{eqnarray}
		Y_{1,3}=&&\frac{8u_1^3}{z^2}-\left(\frac{4u_1+4u_2g+24u_1^2}{z}\right)(\beta\epsilon)\nonumber\\
		&&
		+(2+24u_1)(\beta\epsilon)^2,
	\end{eqnarray}
	
	\begin{eqnarray}
		Y_{1,4}=\left(\frac{12u_1^2}{z}\right)(\beta\epsilon)-(1/2+20u_1)(\beta\epsilon)^2,
	\end{eqnarray}
	
	\begin{eqnarray}
		Y_{1,5}=(6u_1)(\beta\epsilon)^2,
	\end{eqnarray}
\end{subequations}

\begin{subequations}
	\begin{eqnarray}
		Y_{2,0}=\frac{1}{z}+\frac{2+4N_{2e}g}{z^2}-\left(\frac{6}{z}\right)(\beta\epsilon)+(\beta\epsilon)^2,
	\end{eqnarray}
	
	\begin{eqnarray}
		Y_{2,1}=&&-\frac{14u_1+4u_2g}{z^2}+\left(\frac{12+12u_1}{z}\right)(\beta\epsilon)\nonumber\\
		&&
		-6(\beta\epsilon)^2,
	\end{eqnarray}
	
	\begin{eqnarray}
		Y_{2,2}=\frac{12u_1^2}{z^2}-\left(\frac{6+24u_1}{z}\right)(\beta\epsilon)+12(\beta\epsilon)^2,
	\end{eqnarray}
	
	\begin{eqnarray}
		Y_{2,3}=\left(\frac{12u_1}{z}\right)(\beta\epsilon)-10(\beta\epsilon)^2,
	\end{eqnarray}
	
	\begin{eqnarray}
		Y_{2,4}=3(\beta\epsilon)^2,
	\end{eqnarray}
\end{subequations}

\begin{subequations}
	\begin{eqnarray}
		Y_{3,0}=-\frac{22/3}{z^2}+\left(\frac{4}{z}\right)(\beta\epsilon),
	\end{eqnarray}
	
	\begin{eqnarray}
		Y_{3,1}=\frac{8u_1}{z^2}-\left(\frac{8}{z}\right)(\beta\epsilon),
	\end{eqnarray}
	
	\begin{eqnarray}
		Y_{3,2}=\left(\frac{4}{z}\right)(\beta\epsilon),
	\end{eqnarray}
\end{subequations}

\begin{eqnarray}
	Y_{4,0}=\frac{2}{z^2},
\end{eqnarray}

where $N_{3e}$ is defined by half of the number of runs of three consecutive bonds in a single chain, where one of the bonds links a sticker with a non-sticker, and, hence, $N_{3e}=1$ for linear chains.

\bibliography{refs}

\begin{thebibliography}{43}%
\makeatletter
\providecommand \@ifxundefined [1]{%
 \@ifx{#1\undefined}
}%
\providecommand \@ifnum [1]{%
 \ifnum #1\expandafter \@firstoftwo
 \else \expandafter \@secondoftwo
 \fi
}%
\providecommand \@ifx [1]{%
 \ifx #1\expandafter \@firstoftwo
 \else \expandafter \@secondoftwo
 \fi
}%
\providecommand \natexlab [1]{#1}%
\providecommand \enquote  [1]{``#1''}%
\providecommand \bibnamefont  [1]{#1}%
\providecommand \bibfnamefont [1]{#1}%
\providecommand \citenamefont [1]{#1}%
\providecommand \href@noop [0]{\@secondoftwo}%
\providecommand \href [0]{\begingroup \@sanitize@url \@href}%
\providecommand \@href[1]{\@@startlink{#1}\@@href}%
\providecommand \@@href[1]{\endgroup#1\@@endlink}%
\providecommand \@sanitize@url [0]{\catcode `\\12\catcode `\$12\catcode
  `\&12\catcode `\#12\catcode `\^12\catcode `\_12\catcode `\%12\relax}%
\providecommand \@@startlink[1]{}%
\providecommand \@@endlink[0]{}%
\providecommand \url  [0]{\begingroup\@sanitize@url \@url }%
\providecommand \@url [1]{\endgroup\@href {#1}{\urlprefix }}%
\providecommand \urlprefix  [0]{URL }%
\providecommand \Eprint [0]{\href }%
\providecommand \doibase [0]{http://dx.doi.org/}%
\providecommand \selectlanguage [0]{\@gobble}%
\providecommand \bibinfo  [0]{\@secondoftwo}%
\providecommand \bibfield  [0]{\@secondoftwo}%
\providecommand \translation [1]{[#1]}%
\providecommand \BibitemOpen [0]{}%
\providecommand \bibitemStop [0]{}%
\providecommand \bibitemNoStop [0]{.\EOS\space}%
\providecommand \EOS [0]{\spacefactor3000\relax}%
\providecommand \BibitemShut  [1]{\csname bibitem#1\endcsname}%
\let\auto@bib@innerbib\@empty
\bibitem [{\citenamefont {Lo~Verso}\ and\ \citenamefont
  {Likos}(2008)}]{Polymer_49_1425}%
  \BibitemOpen
  \bibfield  {author} {\bibinfo {author} {\bibfnamefont {F.}~\bibnamefont
  {Lo~Verso}}\ and\ \bibinfo {author} {\bibfnamefont {C.~N.}\ \bibnamefont
  {Likos}},\ }\href@noop {} {\bibfield  {journal} {\bibinfo  {journal}
  {Polymer}\ }\textbf {\bibinfo {volume} {49}},\ \bibinfo {pages} {1425}
  (\bibinfo {year} {2008})}\BibitemShut {NoStop}%
\bibitem [{\citenamefont {Athey}(1979)}]{POC_7_289}%
  \BibitemOpen
  \bibfield  {author} {\bibinfo {author} {\bibfnamefont {R.~D.}\ \bibnamefont
  {Athey}, \bibfnamefont {Jr.}},\ }\href@noop {} {\bibfield  {journal}
  {\bibinfo  {journal} {Prog. Org. Coat.}\ }\textbf {\bibinfo {volume} {7}},\
  \bibinfo {pages} {289} (\bibinfo {year} {1979})}\BibitemShut {NoStop}%
\bibitem [{\citenamefont {Goodman}(1989)}]{Book_Goodman}%
  \BibitemOpen
  \bibfield  {author} {\bibinfo {author} {\bibfnamefont {I.}~\bibnamefont
  {Goodman}},\ }in\ \href@noop {} {\emph {\bibinfo {booktitle} {Telechelic
  Polymers: Synthesis and Applications}}},\ \bibinfo {editor} {edited by\
  \bibinfo {editor} {\bibfnamefont {E.~J.}\ \bibnamefont {Goethals}}}\
  (\bibinfo  {publisher} {CRC Press Inc.},\ \bibinfo {address} {Florida},\
  \bibinfo {year} {1989})\ Chap.~\bibinfo {chapter} {I}, p.~\bibinfo {pages}
  {1}\BibitemShut {NoStop}%
\bibitem [{\citenamefont {Kim}\ \emph {et~al.}(2004)\citenamefont {Kim},
  \citenamefont {Kim}, \citenamefont {Kim}, \citenamefont {Jin}, \citenamefont
  {Hong}, \citenamefont {Hwang}, \citenamefont {Cho}, \citenamefont {Shin},\
  and\ \citenamefont {Im}}]{Polymer_45_3527}%
  \BibitemOpen
  \bibfield  {author} {\bibinfo {author} {\bibfnamefont {J.}~\bibnamefont
  {Kim}}, \bibinfo {author} {\bibfnamefont {S.~S.}\ \bibnamefont {Kim}},
  \bibinfo {author} {\bibfnamefont {K.~H.}\ \bibnamefont {Kim}}, \bibinfo
  {author} {\bibfnamefont {Y.~H.}\ \bibnamefont {Jin}}, \bibinfo {author}
  {\bibfnamefont {S.~M.}\ \bibnamefont {Hong}}, \bibinfo {author}
  {\bibfnamefont {S.~S.}\ \bibnamefont {Hwang}}, \bibinfo {author}
  {\bibfnamefont {B.-G.}\ \bibnamefont {Cho}}, \bibinfo {author} {\bibfnamefont
  {D.~Y.}\ \bibnamefont {Shin}}, \ and\ \bibinfo {author} {\bibfnamefont
  {S.~S.}\ \bibnamefont {Im}},\ }\href@noop {} {\bibfield  {journal} {\bibinfo
  {journal} {Polymer}\ }\textbf {\bibinfo {volume} {45}},\ \bibinfo {pages}
  {3527} (\bibinfo {year} {2004})}\BibitemShut {NoStop}%
\bibitem [{\citenamefont {de~Greef}\ and\ \citenamefont
  {Meijer}(2008)}]{Nature_453_171}%
  \BibitemOpen
  \bibfield  {author} {\bibinfo {author} {\bibfnamefont {T.~F.~A.}\
  \bibnamefont {de~Greef}}\ and\ \bibinfo {author} {\bibfnamefont {E.~W.}\
  \bibnamefont {Meijer}},\ }\href@noop {} {\bibfield  {journal} {\bibinfo
  {journal} {Nature (London)}\ }\textbf {\bibinfo {volume} {453}},\ \bibinfo
  {pages} {171} (\bibinfo {year} {2008})}\BibitemShut {NoStop}%
\bibitem [{\citenamefont {Semenov}, \citenamefont {Joanny},\ and\ \citenamefont
  {Khokhlov}(1995)}]{Mac_28_1066}%
  \BibitemOpen
  \bibfield  {author} {\bibinfo {author} {\bibfnamefont {A.~N.}\ \bibnamefont
  {Semenov}}, \bibinfo {author} {\bibfnamefont {J.-F.}\ \bibnamefont {Joanny}},
  \ and\ \bibinfo {author} {\bibfnamefont {A.~R.}\ \bibnamefont {Khokhlov}},\
  }\href@noop {} {\bibfield  {journal} {\bibinfo  {journal} {Macromolecules}\
  }\textbf {\bibinfo {volume} {28}},\ \bibinfo {pages} {1066} (\bibinfo {year}
  {1995})}\BibitemShut {NoStop}%
\bibitem [{\citenamefont {Semenov}, \citenamefont {Nyrkova},\ and\
  \citenamefont {Cates}(1995)}]{Mac_28_7879}%
  \BibitemOpen
  \bibfield  {author} {\bibinfo {author} {\bibfnamefont {A.~N.}\ \bibnamefont
  {Semenov}}, \bibinfo {author} {\bibfnamefont {I.~A.}\ \bibnamefont
  {Nyrkova}}, \ and\ \bibinfo {author} {\bibfnamefont {M.~E.}\ \bibnamefont
  {Cates}},\ }\href@noop {} {\bibfield  {journal} {\bibinfo  {journal}
  {Macromolecules}\ }\textbf {\bibinfo {volume} {28}},\ \bibinfo {pages} {7879}
  (\bibinfo {year} {1995})}\BibitemShut {NoStop}%
\bibitem [{\citenamefont {Ermoshkin}\ and\ \citenamefont
  {Erukhimovich}(1999)}]{JCP_110_1781}%
  \BibitemOpen
  \bibfield  {author} {\bibinfo {author} {\bibfnamefont {A.~V.}\ \bibnamefont
  {Ermoshkin}}\ and\ \bibinfo {author} {\bibfnamefont {I.}~\bibnamefont
  {Erukhimovich}},\ }\href@noop {} {\bibfield  {journal} {\bibinfo  {journal}
  {J. Chem. Phys.}\ }\textbf {\bibinfo {volume} {110}},\ \bibinfo {pages}
  {1781} (\bibinfo {year} {1999})}\BibitemShut {NoStop}%
\bibitem [{\citenamefont {Kolbet}\ and\ \citenamefont
  {Schweizer}(2000{\natexlab{a}})}]{Mac_33_1425}%
  \BibitemOpen
  \bibfield  {author} {\bibinfo {author} {\bibfnamefont {K.~A.}\ \bibnamefont
  {Kolbet}}\ and\ \bibinfo {author} {\bibfnamefont {K.~S.}\ \bibnamefont
  {Schweizer}},\ }\href@noop {} {\bibfield  {journal} {\bibinfo  {journal}
  {Macromolecules}\ }\textbf {\bibinfo {volume} {33}},\ \bibinfo {pages} {1425}
  (\bibinfo {year} {2000}{\natexlab{a}})}\BibitemShut {NoStop}%
\bibitem [{\citenamefont {Kolbet}\ and\ \citenamefont
  {Schweizer}(2000{\natexlab{b}})}]{Mac_33_1443}%
  \BibitemOpen
  \bibfield  {author} {\bibinfo {author} {\bibfnamefont {K.~A.}\ \bibnamefont
  {Kolbet}}\ and\ \bibinfo {author} {\bibfnamefont {K.~S.}\ \bibnamefont
  {Schweizer}},\ }\href@noop {} {\bibfield  {journal} {\bibinfo  {journal}
  {Macromolecules}\ }\textbf {\bibinfo {volume} {33}},\ \bibinfo {pages} {1443}
  (\bibinfo {year} {2000}{\natexlab{b}})}\BibitemShut {NoStop}%
\bibitem [{\citenamefont {Sung}\ and\ \citenamefont
  {Yethiraj}(2003)}]{JCP_119_6916}%
  \BibitemOpen
  \bibfield  {author} {\bibinfo {author} {\bibfnamefont {B.~J.}\ \bibnamefont
  {Sung}}\ and\ \bibinfo {author} {\bibfnamefont {A.}~\bibnamefont
  {Yethiraj}},\ }\href@noop {} {\bibfield  {journal} {\bibinfo  {journal} {J.
  Chem. Phys.}\ }\textbf {\bibinfo {volume} {119}},\ \bibinfo {pages} {6916}
  (\bibinfo {year} {2003})}\BibitemShut {NoStop}%
\bibitem [{\citenamefont {Bohbot-Raviv}, \citenamefont {Snyder},\ and\
  \citenamefont {Wang}(2004)}]{Lan_20_7860}%
  \BibitemOpen
  \bibfield  {author} {\bibinfo {author} {\bibfnamefont {Y.}~\bibnamefont
  {Bohbot-Raviv}}, \bibinfo {author} {\bibfnamefont {T.~M.}\ \bibnamefont
  {Snyder}}, \ and\ \bibinfo {author} {\bibfnamefont {Z.-G.}\ \bibnamefont
  {Wang}},\ }\href@noop {} {\bibfield  {journal} {\bibinfo  {journal}
  {Langmuir}\ }\textbf {\bibinfo {volume} {20}},\ \bibinfo {pages} {7860}
  (\bibinfo {year} {2004})}\BibitemShut {NoStop}%
\bibitem [{\citenamefont {Anthamatten}(2007)}]{JPSB_45_3285}%
  \BibitemOpen
  \bibfield  {author} {\bibinfo {author} {\bibfnamefont {M.}~\bibnamefont
  {Anthamatten}},\ }\href@noop {} {\bibfield  {journal} {\bibinfo  {journal}
  {J. Polym. Sci., Part B: Polym. Phys.}\ }\textbf {\bibinfo {volume} {45}},\
  \bibinfo {pages} {3285} (\bibinfo {year} {2007})}\BibitemShut {NoStop}%
\bibitem [{\citenamefont {Elliott}\ and\ \citenamefont
  {Fredrickson}(2009)}]{JCP_131_144906}%
  \BibitemOpen
  \bibfield  {author} {\bibinfo {author} {\bibfnamefont {R.}~\bibnamefont
  {Elliott}}\ and\ \bibinfo {author} {\bibfnamefont {G.~H.}\ \bibnamefont
  {Fredrickson}},\ }\href@noop {} {\bibfield  {journal} {\bibinfo  {journal}
  {J. Chem. Phys.}\ }\textbf {\bibinfo {volume} {131}},\ \bibinfo {pages}
  {144906} (\bibinfo {year} {2009})}\BibitemShut {NoStop}%
\bibitem [{\citenamefont {Bymaster}\ and\ \citenamefont
  {Chapman}(2010)}]{JPCB_114_12298}%
  \BibitemOpen
  \bibfield  {author} {\bibinfo {author} {\bibfnamefont {A.}~\bibnamefont
  {Bymaster}}\ and\ \bibinfo {author} {\bibfnamefont {W.~G.}\ \bibnamefont
  {Chapman}},\ }\href@noop {} {\bibfield  {journal} {\bibinfo  {journal} {J.
  Phys. Chem. B}\ }\textbf {\bibinfo {volume} {114}},\ \bibinfo {pages} {12298}
  (\bibinfo {year} {2010})}\BibitemShut {NoStop}%
\bibitem [{\citenamefont {Balazs}, \citenamefont {Anderson},\ and\
  \citenamefont {Muthukumar}(1987)}]{Mac_20_1999}%
  \BibitemOpen
  \bibfield  {author} {\bibinfo {author} {\bibfnamefont {A.~C.}\ \bibnamefont
  {Balazs}}, \bibinfo {author} {\bibfnamefont {C.}~\bibnamefont {Anderson}}, \
  and\ \bibinfo {author} {\bibfnamefont {M.}~\bibnamefont {Muthukumar}},\
  }\href@noop {} {\bibfield  {journal} {\bibinfo  {journal} {Macromolecules}\
  }\textbf {\bibinfo {volume} {20}},\ \bibinfo {pages} {1999} (\bibinfo {year}
  {1987})}\BibitemShut {NoStop}%
\bibitem [{\citenamefont {Khalatur}\ \emph {et~al.}(1999)\citenamefont
  {Khalatur}, \citenamefont {Khokhlov}, \citenamefont {Kovalenko},\ and\
  \citenamefont {Mologin}}]{JCP_110_6039}%
  \BibitemOpen
  \bibfield  {author} {\bibinfo {author} {\bibfnamefont {P.~G.}\ \bibnamefont
  {Khalatur}}, \bibinfo {author} {\bibfnamefont {A.~R.}\ \bibnamefont
  {Khokhlov}}, \bibinfo {author} {\bibfnamefont {J.~N.}\ \bibnamefont
  {Kovalenko}}, \ and\ \bibinfo {author} {\bibfnamefont {D.~A.}\ \bibnamefont
  {Mologin}},\ }\href@noop {} {\bibfield  {journal} {\bibinfo  {journal} {J.
  Chem. Phys.}\ }\textbf {\bibinfo {volume} {110}},\ \bibinfo {pages} {6039}
  (\bibinfo {year} {1999})}\BibitemShut {NoStop}%
\bibitem [{\citenamefont {Bedrov}, \citenamefont {Smith},\ and\ \citenamefont
  {Douglas}(2002)}]{EPL_59_384}%
  \BibitemOpen
  \bibfield  {author} {\bibinfo {author} {\bibfnamefont {D.}~\bibnamefont
  {Bedrov}}, \bibinfo {author} {\bibfnamefont {G.~D.}\ \bibnamefont {Smith}}, \
  and\ \bibinfo {author} {\bibfnamefont {J.~F.}\ \bibnamefont {Douglas}},\
  }\href@noop {} {\bibfield  {journal} {\bibinfo  {journal} {Europhys. Lett.}\
  }\textbf {\bibinfo {volume} {59}},\ \bibinfo {pages} {384} (\bibinfo {year}
  {2002})}\BibitemShut {NoStop}%
\bibitem [{\citenamefont {Bedrov}, \citenamefont {Smith},\ and\ \citenamefont
  {Douglas}(2004)}]{Polymer_45_3961}%
  \BibitemOpen
  \bibfield  {author} {\bibinfo {author} {\bibfnamefont {D.}~\bibnamefont
  {Bedrov}}, \bibinfo {author} {\bibfnamefont {G.~D.}\ \bibnamefont {Smith}}, \
  and\ \bibinfo {author} {\bibfnamefont {J.~F.}\ \bibnamefont {Douglas}},\
  }\href@noop {} {\bibfield  {journal} {\bibinfo  {journal} {Polymer}\ }\textbf
  {\bibinfo {volume} {45}},\ \bibinfo {pages} {3961} (\bibinfo {year}
  {2004})}\BibitemShut {NoStop}%
\bibitem [{\citenamefont {Loverde}, \citenamefont {Ermoshkin},\ and\
  \citenamefont {Olvera de~la Cruz}(2005)}]{JPSB_43_796}%
  \BibitemOpen
  \bibfield  {author} {\bibinfo {author} {\bibfnamefont {S.~M.}\ \bibnamefont
  {Loverde}}, \bibinfo {author} {\bibfnamefont {A.~V.}\ \bibnamefont
  {Ermoshkin}}, \ and\ \bibinfo {author} {\bibfnamefont {M.}~\bibnamefont
  {Olvera de~la Cruz}},\ }\href@noop {} {\bibfield  {journal} {\bibinfo
  {journal} {J. Polym. Sci., Part B: Polym. Phys.}\ }\textbf {\bibinfo {volume}
  {43}},\ \bibinfo {pages} {796} (\bibinfo {year} {2005})}\BibitemShut
  {NoStop}%
\bibitem [{\citenamefont {Lo~Verso}\ \emph {et~al.}(2006)\citenamefont
  {Lo~Verso}, \citenamefont {Likos}, \citenamefont {Mayer},\ and\ \citenamefont
  {L{\"{o}}wen}}]{PRL_96_187802}%
  \BibitemOpen
  \bibfield  {author} {\bibinfo {author} {\bibfnamefont {F.}~\bibnamefont
  {Lo~Verso}}, \bibinfo {author} {\bibfnamefont {C.~N.}\ \bibnamefont {Likos}},
  \bibinfo {author} {\bibfnamefont {C.}~\bibnamefont {Mayer}}, \ and\ \bibinfo
  {author} {\bibfnamefont {H.}~\bibnamefont {L{\"{o}}wen}},\ }\href@noop {}
  {\bibfield  {journal} {\bibinfo  {journal} {Phys. Rev. Lett.}\ }\textbf
  {\bibinfo {volume} {96}},\ \bibinfo {pages} {187802} (\bibinfo {year}
  {2006})}\BibitemShut {NoStop}%
\bibitem [{\citenamefont {Capone}\ \emph {et~al.}(2012)\citenamefont {Capone},
  \citenamefont {Coluzza}, \citenamefont {Lo~Verso}, \citenamefont {Likos},\
  and\ \citenamefont {Blaak}}]{PRL_109_238301}%
  \BibitemOpen
  \bibfield  {author} {\bibinfo {author} {\bibfnamefont {B.}~\bibnamefont
  {Capone}}, \bibinfo {author} {\bibfnamefont {I.}~\bibnamefont {Coluzza}},
  \bibinfo {author} {\bibfnamefont {F.}~\bibnamefont {Lo~Verso}}, \bibinfo
  {author} {\bibfnamefont {C.~N.}\ \bibnamefont {Likos}}, \ and\ \bibinfo
  {author} {\bibfnamefont {R.}~\bibnamefont {Blaak}},\ }\href@noop {}
  {\bibfield  {journal} {\bibinfo  {journal} {Phys. Rev. Lett.}\ }\textbf
  {\bibinfo {volume} {109}},\ \bibinfo {pages} {238301} (\bibinfo {year}
  {2012})}\BibitemShut {NoStop}%
\bibitem [{\citenamefont {Baljon}, \citenamefont {Flynn},\ and\ \citenamefont
  {Krawzsenek}(2007)}]{JCP_126_044907}%
  \BibitemOpen
  \bibfield  {author} {\bibinfo {author} {\bibfnamefont {A.~R.~C.}\
  \bibnamefont {Baljon}}, \bibinfo {author} {\bibfnamefont {D.}~\bibnamefont
  {Flynn}}, \ and\ \bibinfo {author} {\bibfnamefont {D.}~\bibnamefont
  {Krawzsenek}},\ }\href@noop {} {\bibfield  {journal} {\bibinfo  {journal} {J.
  Chem. Phys.}\ }\textbf {\bibinfo {volume} {126}},\ \bibinfo {pages} {044907}
  (\bibinfo {year} {2007})}\BibitemShut {NoStop}%
\bibitem [{\citenamefont {Cass}\ \emph {et~al.}(2008)\citenamefont {Cass},
  \citenamefont {Heyes}, \citenamefont {Blanchard},\ and\ \citenamefont
  {English}}]{JPCM_20_335103}%
  \BibitemOpen
  \bibfield  {author} {\bibinfo {author} {\bibfnamefont {M.~J.}\ \bibnamefont
  {Cass}}, \bibinfo {author} {\bibfnamefont {D.~M.}\ \bibnamefont {Heyes}},
  \bibinfo {author} {\bibfnamefont {R.-L.}\ \bibnamefont {Blanchard}}, \ and\
  \bibinfo {author} {\bibfnamefont {R.~J.}\ \bibnamefont {English}},\
  }\href@noop {} {\bibfield  {journal} {\bibinfo  {journal} {J. Phys.: Condens.
  Matter}\ }\textbf {\bibinfo {volume} {20}},\ \bibinfo {pages} {335103}
  (\bibinfo {year} {2008})}\BibitemShut {NoStop}%
\bibitem [{\citenamefont {Myung}\ \emph {et~al.}(2014)\citenamefont {Myung},
  \citenamefont {Taslimi}, \citenamefont {Winkler},\ and\ \citenamefont
  {Gompper}}]{Mac_47_4118}%
  \BibitemOpen
  \bibfield  {author} {\bibinfo {author} {\bibfnamefont {J.~S.}\ \bibnamefont
  {Myung}}, \bibinfo {author} {\bibfnamefont {F.}~\bibnamefont {Taslimi}},
  \bibinfo {author} {\bibfnamefont {R.~G.}\ \bibnamefont {Winkler}}, \ and\
  \bibinfo {author} {\bibfnamefont {G.}~\bibnamefont {Gompper}},\ }\href@noop
  {} {\bibfield  {journal} {\bibinfo  {journal} {Macromolecules}\ }\textbf
  {\bibinfo {volume} {47}},\ \bibinfo {pages} {4118} (\bibinfo {year}
  {2014})}\BibitemShut {NoStop}%
\bibitem [{\citenamefont {Taslimi}, \citenamefont {Gompper},\ and\
  \citenamefont {Winkler}(2014)}]{Mac_47_6946}%
  \BibitemOpen
  \bibfield  {author} {\bibinfo {author} {\bibfnamefont {F.}~\bibnamefont
  {Taslimi}}, \bibinfo {author} {\bibfnamefont {G.}~\bibnamefont {Gompper}}, \
  and\ \bibinfo {author} {\bibfnamefont {R.~G.}\ \bibnamefont {Winkler}},\
  }\href@noop {} {\bibfield  {journal} {\bibinfo  {journal} {Macromolecules}\
  }\textbf {\bibinfo {volume} {47}},\ \bibinfo {pages} {6946} (\bibinfo {year}
  {2014})}\BibitemShut {NoStop}%
\bibitem [{\citenamefont {Dudowicz}\ and\ \citenamefont
  {Freed}(2012)}]{JCP_136_064902}%
  \BibitemOpen
  \bibfield  {author} {\bibinfo {author} {\bibfnamefont {J.}~\bibnamefont
  {Dudowicz}}\ and\ \bibinfo {author} {\bibfnamefont {K.~F.}\ \bibnamefont
  {Freed}},\ }\href@noop {} {\bibfield  {journal} {\bibinfo  {journal} {J.
  Chem. Phys.}\ }\textbf {\bibinfo {volume} {136}},\ \bibinfo {pages} {064902}
  (\bibinfo {year} {2012})}\BibitemShut {NoStop}%
\bibitem [{\citenamefont {Nemirovsky}, \citenamefont {Bawendi},\ and\
  \citenamefont {Freed}(1987)}]{JCP_87_7272}%
  \BibitemOpen
  \bibfield  {author} {\bibinfo {author} {\bibfnamefont {A.~M.}\ \bibnamefont
  {Nemirovsky}}, \bibinfo {author} {\bibfnamefont {M.~G.}\ \bibnamefont
  {Bawendi}}, \ and\ \bibinfo {author} {\bibfnamefont {K.~F.}\ \bibnamefont
  {Freed}},\ }\href@noop {} {\bibfield  {journal} {\bibinfo  {journal} {J.
  Chem. Phys.}\ }\textbf {\bibinfo {volume} {87}},\ \bibinfo {pages} {7272}
  (\bibinfo {year} {1987})}\BibitemShut {NoStop}%
\bibitem [{\citenamefont {Dudowicz}\ and\ \citenamefont
  {Freed}(1991)}]{Mac_24_5076}%
  \BibitemOpen
  \bibfield  {author} {\bibinfo {author} {\bibfnamefont {J.}~\bibnamefont
  {Dudowicz}}\ and\ \bibinfo {author} {\bibfnamefont {K.~F.}\ \bibnamefont
  {Freed}},\ }\href@noop {} {\bibfield  {journal} {\bibinfo  {journal}
  {Macromolecules}\ }\textbf {\bibinfo {volume} {24}},\ \bibinfo {pages} {5076}
  (\bibinfo {year} {1991})}\BibitemShut {NoStop}%
\bibitem [{\citenamefont {Foreman}\ and\ \citenamefont
  {Freed}(1998)}]{ACP_103_335}%
  \BibitemOpen
  \bibfield  {author} {\bibinfo {author} {\bibfnamefont {K.~W.}\ \bibnamefont
  {Foreman}}\ and\ \bibinfo {author} {\bibfnamefont {K.~F.}\ \bibnamefont
  {Freed}},\ }\href@noop {} {\bibfield  {journal} {\bibinfo  {journal} {Adv.
  Chem. Phys.}\ }\textbf {\bibinfo {volume} {103}},\ \bibinfo {pages} {335}
  (\bibinfo {year} {1998})}\BibitemShut {NoStop}%
\bibitem [{\citenamefont {Freed}\ and\ \citenamefont
  {Dudowicz}(2005)}]{APS_183_63}%
  \BibitemOpen
  \bibfield  {author} {\bibinfo {author} {\bibfnamefont {K.~F.}\ \bibnamefont
  {Freed}}\ and\ \bibinfo {author} {\bibfnamefont {J.}~\bibnamefont
  {Dudowicz}},\ }\href@noop {} {\bibfield  {journal} {\bibinfo  {journal} {Adv.
  Polym. Sci.}\ }\textbf {\bibinfo {volume} {183}},\ \bibinfo {pages} {63}
  (\bibinfo {year} {2005})}\BibitemShut {NoStop}%
\bibitem [{\citenamefont {Dudowicz}, \citenamefont {Freed},\ and\ \citenamefont
  {Douglas}(2012{\natexlab{a}})}]{JCP_136_064903}%
  \BibitemOpen
  \bibfield  {author} {\bibinfo {author} {\bibfnamefont {J.}~\bibnamefont
  {Dudowicz}}, \bibinfo {author} {\bibfnamefont {K.~F.}\ \bibnamefont {Freed}},
  \ and\ \bibinfo {author} {\bibfnamefont {J.~F.}\ \bibnamefont {Douglas}},\
  }\href@noop {} {\bibfield  {journal} {\bibinfo  {journal} {J. Chem. Phys.}\
  }\textbf {\bibinfo {volume} {136}},\ \bibinfo {pages} {064903} (\bibinfo
  {year} {2012}{\natexlab{a}})}\BibitemShut {NoStop}%
\bibitem [{\citenamefont {Dudowicz}, \citenamefont {Freed},\ and\ \citenamefont
  {Douglas}(2012{\natexlab{b}})}]{JCP_136_194902}%
  \BibitemOpen
  \bibfield  {author} {\bibinfo {author} {\bibfnamefont {J.}~\bibnamefont
  {Dudowicz}}, \bibinfo {author} {\bibfnamefont {K.~F.}\ \bibnamefont {Freed}},
  \ and\ \bibinfo {author} {\bibfnamefont {J.~F.}\ \bibnamefont {Douglas}},\
  }\href@noop {} {\bibfield  {journal} {\bibinfo  {journal} {J. Chem. Phys.}\
  }\textbf {\bibinfo {volume} {136}},\ \bibinfo {pages} {194902} (\bibinfo
  {year} {2012}{\natexlab{b}})}\BibitemShut {NoStop}%
\bibitem [{\citenamefont {Dudowicz}, \citenamefont {Freed},\ and\ \citenamefont
  {Douglas}(2008)}]{ACP_137_125}%
  \BibitemOpen
  \bibfield  {author} {\bibinfo {author} {\bibfnamefont {J.}~\bibnamefont
  {Dudowicz}}, \bibinfo {author} {\bibfnamefont {K.~F.}\ \bibnamefont {Freed}},
  \ and\ \bibinfo {author} {\bibfnamefont {J.~F.}\ \bibnamefont {Douglas}},\
  }\href@noop {} {\bibfield  {journal} {\bibinfo  {journal} {Adv. Chem. Phys.}\
  }\textbf {\bibinfo {volume} {137}},\ \bibinfo {pages} {125} (\bibinfo {year}
  {2008})}\BibitemShut {NoStop}%
\bibitem [{\citenamefont {Adam}\ and\ \citenamefont
  {Gibbs}(1965)}]{JCP_43_139}%
  \BibitemOpen
  \bibfield  {author} {\bibinfo {author} {\bibfnamefont {G.}~\bibnamefont
  {Adam}}\ and\ \bibinfo {author} {\bibfnamefont {J.~H.}\ \bibnamefont
  {Gibbs}},\ }\href@noop {} {\bibfield  {journal} {\bibinfo  {journal} {J.
  Chem. Phys.}\ }\textbf {\bibinfo {volume} {43}},\ \bibinfo {pages} {139}
  (\bibinfo {year} {1965})}\BibitemShut {NoStop}%
\bibitem [{\citenamefont {Freed}(2014)}]{JCP_141_141102}%
  \BibitemOpen
  \bibfield  {author} {\bibinfo {author} {\bibfnamefont {K.~F.}\ \bibnamefont
  {Freed}},\ }\href@noop {} {\bibfield  {journal} {\bibinfo  {journal} {J.
  Chem. Phys.}\ }\textbf {\bibinfo {volume} {141}},\ \bibinfo {pages} {141102}
  (\bibinfo {year} {2014})}\BibitemShut {NoStop}%
\bibitem [{\citenamefont {Dudowicz}, \citenamefont {Douglas},\ and\
  \citenamefont {Freed}(2014)}]{JCP_140_244905}%
  \BibitemOpen
  \bibfield  {author} {\bibinfo {author} {\bibfnamefont {J.}~\bibnamefont
  {Dudowicz}}, \bibinfo {author} {\bibfnamefont {J.~F.}\ \bibnamefont
  {Douglas}}, \ and\ \bibinfo {author} {\bibfnamefont {K.~F.}\ \bibnamefont
  {Freed}},\ }\href@noop {} {\bibfield  {journal} {\bibinfo  {journal} {J.
  Chem. Phys.}\ }\textbf {\bibinfo {volume} {140}},\ \bibinfo {pages} {244905}
  (\bibinfo {year} {2014})}\BibitemShut {NoStop}%
\bibitem [{\citenamefont {Xu}\ and\ \citenamefont
  {Freed}(2014)}]{JCP_141_044909}%
  \BibitemOpen
  \bibfield  {author} {\bibinfo {author} {\bibfnamefont {W.-S.}\ \bibnamefont
  {Xu}}\ and\ \bibinfo {author} {\bibfnamefont {K.~F.}\ \bibnamefont {Freed}},\
  }\href@noop {} {\bibfield  {journal} {\bibinfo  {journal} {J. Chem. Phys.}\
  }\textbf {\bibinfo {volume} {141}},\ \bibinfo {pages} {044909} (\bibinfo
  {year} {2014})}\BibitemShut {NoStop}%
\bibitem [{\citenamefont {Freed}(2012)}]{JCP_136_244904}%
  \BibitemOpen
  \bibfield  {author} {\bibinfo {author} {\bibfnamefont {K.~F.}\ \bibnamefont
  {Freed}},\ }\href@noop {} {\bibfield  {journal} {\bibinfo  {journal} {J.
  Chem. Phys.}\ }\textbf {\bibinfo {volume} {136}},\ \bibinfo {pages} {244904}
  (\bibinfo {year} {2012})}\BibitemShut {NoStop}%
\bibitem [{\citenamefont {Jacobson}\ and\ \citenamefont
  {Stockmayer}(1950)}]{JCP_18_1600}%
  \BibitemOpen
  \bibfield  {author} {\bibinfo {author} {\bibfnamefont {H.}~\bibnamefont
  {Jacobson}}\ and\ \bibinfo {author} {\bibfnamefont {W.~H.}\ \bibnamefont
  {Stockmayer}},\ }\href@noop {} {\bibfield  {journal} {\bibinfo  {journal} {J.
  Chem. Phys.}\ }\textbf {\bibinfo {volume} {18}},\ \bibinfo {pages} {1600}
  (\bibinfo {year} {1950})}\BibitemShut {NoStop}%
\bibitem [{\citenamefont {Bawendi}, \citenamefont {Freed},\ and\ \citenamefont
  {Mohanty}(1987)}]{JCP_87_5534}%
  \BibitemOpen
  \bibfield  {author} {\bibinfo {author} {\bibfnamefont {M.~G.}\ \bibnamefont
  {Bawendi}}, \bibinfo {author} {\bibfnamefont {K.~F.}\ \bibnamefont {Freed}},
  \ and\ \bibinfo {author} {\bibfnamefont {U.}~\bibnamefont {Mohanty}},\
  }\href@noop {} {\bibfield  {journal} {\bibinfo  {journal} {J. Chem. Phys.}\
  }\textbf {\bibinfo {volume} {87}},\ \bibinfo {pages} {5534} (\bibinfo {year}
  {1987})}\BibitemShut {NoStop}%
\bibitem [{\citenamefont {Freed}(2009)}]{JCP_130_061103}%
  \BibitemOpen
  \bibfield  {author} {\bibinfo {author} {\bibfnamefont {K.~F.}\ \bibnamefont
  {Freed}},\ }\href@noop {} {\bibfield  {journal} {\bibinfo  {journal} {J.
  Chem. Phys.}\ }\textbf {\bibinfo {volume} {130}},\ \bibinfo {pages} {061103}
  (\bibinfo {year} {2009})}\BibitemShut {NoStop}%
\bibitem [{\citenamefont {Xu}\ and\ \citenamefont {Freed}()}]{Paper2}%
  \BibitemOpen
  \bibfield  {author} {\bibinfo {author} {\bibfnamefont {W.-S.}\ \bibnamefont
  {Xu}}\ and\ \bibinfo {author} {\bibfnamefont {K.~F.}\ \bibnamefont {Freed}},\
  }\href@noop {} {\enquote {\bibinfo {title} {Lattice model of linear
  telechelic polymer melts. {II}. {I}nfluence of chain stiffness on basic
  thermodynamic properties},}\ }\bibinfo {note} {{J. Chem. Phys.}
  (accepted)}\BibitemShut {NoStop}%
\end{thebibliography}%

\end{document}